\documentstyle[emulateapj]{article}
\newcommand{\op}{\rm Ly$\alpha$\ }
\newcommand{\msun}{\mbox{M$_\odot$}}
\newcommand{\kms}{\, {\rm km\, s}^{-1}}
\newcommand{\cm}{\, {\rm cm}}
\newcommand{\s}{\, {\rm s}}
\newcommand{\Hz}{\, {\rm Hz}}
\newcommand{\sr}{\, {\rm sr}}
\newcommand{\erg}{\, {\rm ergs}}
\newcommand{\kpc}{\, {\rm kpc}}
\newcommand{\mpc}{\, {\rm Mpc}}

\newcommand{\lya}{Ly$\alpha$ }
\newcommand{\etal}{et al.\ }
\newcommand{\yr}{\, {\rm yr}}

\newcommand{\ergs}{\mbox{ergs\,s$^{-1}$}}
\newcommand{\K}{\, {\rm K}}

\newcommand{\hi}{\mbox{H\,{\scriptsize I}\ }}

\newcommand{\heii}{\mbox{He\,{\scriptsize II}\ }}
\newcommand{\heiii}{\mbox{He\,{\scriptsize III}\ }}
\newcommand{\CIV}{\mbox{C\,{\scriptsize IV}}}

\newcommand{\SiIV}{\mbox{Si\,{\scriptsize IV}}}

\def\spose#1{\hbox to 0pt{#1\hss}}
\def\lta{\mathrel{\spose{\lower 3pt\hbox{$\mathchar"218$}}
     \raise 2.0pt\hbox{$\mathchar"13C$}}}
\def\gta{\mathrel{\spose{\lower 3pt\hbox{$\mathchar"218$}}
     \raise 2.0pt\hbox{$\mathchar"13E$}}}

\makeatletter

\newenvironment{figurehere}
  {\def\@captype{figure}}
  {}
\makeatother

\begin{document}
\textheight=24.5cm
\hoffset=-1.75cm

%
\title{An ionizing UV background dominated by massive stars}

\lefthead{Haehnelt, Madau, Kudritzki, \& Haardt}
\righthead{An ionizing UV background dominated by 
massive stars}

\author{Martin G. Haehnelt\altaffilmark{1,2},
Piero Madau\altaffilmark{3,4}, Rolf Kudritzki\altaffilmark{5},
and Francesco Haardt\altaffilmark{6}}
\altaffiltext{1}{Max-Planck-Institut f\"ur Astrophysik, 
Karl-Schwarzschild-Str.1, Postfach 1317, D-85741 Garching, Germany.}
\altaffiltext{2}{Astrophysics Group, Imperial College of Science, 
Technology and Medicine, Prince Consort Road, London SW7 2BW, UK.}
\altaffiltext{3}{Institute of Astronomy, University of Cambridge,
Madingley Road, Cambridge CB3 0HA, UK.} 
\altaffiltext{4}{Department of Astronomy and Astrophysics, University of
California Santa Cruz, Santa Cruz, CA 95064.}
\altaffiltext{5}{Institute for Astronomy, University of Hawaii, 2680
Woodlawn Drive, Honolulu, HI 96822.} 
\altaffiltext{6}{Dipartimento di Scienze, Universit\'a dell'Insubria,
via Lucini 3, Como, Italy.}                                


\begin{abstract}
We discuss the implications of a stellar-dominated UV background at 
high redshifts for the star formation history of Lyman-break galaxies (LBGs) 
and the thermal and ionization state of the intergalactic
medium (IGM). The composite spectrum of 29 LBGs evaluated
by Steidel \etal (2000) at $\langle z\rangle=3.4$ can be well fit by 
a stellar population with ongoing star formation, a Salpeter initial 
mass function, modest or negligible dust reddening, and 
no intrinsic \hi photoelectric absorption.
Fading starbursts in which star formation has ceased  
for $10^{7}\,$ yr or more cannot reproduce the 
observed flux shortward of 1 Ryd. 
The small \hi optical depth in LBGs suggests that the neutral gas from 
which stars form is most likely contained in compact 
clouds of neutral gas with small covering factor.
The escape fraction of H-ionizing photons  must
be close to 100 percent for the observed sample of LBGs.
The spectrum of ionizing photons  produced by 
a stellar population with ongoing star formation 
is similar to that of QSOs between 1 and 3 Ryd,
but becomes softer between 3 and 4 Ryd and drops sharply shortward   
of 4 Ryd. A galaxy-dominated UV background appears inconsistent with the 
observed \heii/\hi opacity ratio at $z=2.4$, but might be able to 
explain the \SiIV/\CIV\ abundances measured at $z>3$ in quasar 
absorption spectra. 
A scenario may be emerging where star-forming galaxies reionized 
intergalactic hydrogen at $z>6$ and dominate the 1 Ryd metagalactic flux 
at $z>3$, with quasi-stellar sources taking over at lower redshifts.
If the large amplitude of the H-ionizing flux estimated 
by Steidel \etal is correct,  hydrodynamical simulations of structure
formation in the IGM within the cold dark matter paradigm require
a baryon density (to explain the observed opacity of the \op forest
in quasar absorption spectra) which is similar to or larger than that 
favoured by recent CMB experiments, and is inconsistent with standard 
nucleosynthesis values.

\end{abstract}


\section{Introduction}
The strength and spectrum of the UV ionizing background at high redshift 
and the nature 
of the sources which reionized the hydrogen component of the intergalactic
medium (IGM) are two of the big outstanding questions in observational 
cosmology. QSOs and star-forming galaxies have long been considered 
the two prime candidates.
It has been argued that QSOs may fall short of producing a metagalactic 
flux as high as that inferred from the proximity effect  at high redshift
(Bechtold 1994; Giallongo \etal 1996; Cooke \etal 1997; Scott \etal 2000) 
since their space density declines rapidly at early epochs (Shapiro \& 
Giroux 1987; Miralda-Escud\'e \& Ostriker 1990; Meiksin \& Madau 
1993; Madau \etal 1999). 
The detection of a numerous population of Lyman-break galaxies (LBGs)
at $z\approx 3$ (Steidel \etal 1996) makes the idea of massive stars 
dominating the UV background at early epochs quite plausible. 
At $1500\,$\AA\ their emissivity outweighs that of QSOs by 
about a factor of fifteen (Madau \etal
1999). Until recently it seemed likely that only a small 
fraction of the ionizing photons emitted within
starburst galaxies could escape the dense \hi layers into the IGM, 
as is observed in their low-redshift counterparts (Leitherer
\etal 1995; Hurwitz \etal 1997). 
The detection by Steidel \etal (2000) of flux beyond the Lyman 
limit (with no significant indication of a break at $912\,$\AA) in a composite 
spectrum of 
29 LBGs at $\langle z\rangle=3.4$ thus comes as a surprise. 

In this {\it Letter} we assess the significance of the Steidel \etal (2000) 
findings by comparing their spectrum to model spectra  
calculated with the population synthesis code STARBURST99 (Leitherer \etal 
1999). We discuss the implications of this result for the star formation 
history and gas content of these galaxies, for the spectrum of the 
UV background, for the cosmological reionization, and 
for the baryon content of the Universe.

\section{Implications for the star formation history} 

The most striking feature of the composite spectrum of Steidel \etal  
is the apparent lack of a significant drop at the Lyman limit. 
Stars with effective temperatures  $\lta 30,000\,$K 
have a strong intrinsic break
at the Lyman edge. The composite  spectrum shortwards of 1 Ryd 
is thus dominated by 
O stars with masses $\ga 20 \msun$  and lifetimes $\la 10^7\yr$.
Note that the signatures of O stars have already been detected at longer 
wavelengths (Pettini \etal 2000) so their presence does not come as a 
surprise. Considering the  large space density and luminosities of LBGs 
(Steidel \etal 1996)  it seems very unlikely that these galaxies have
been caught during a very short starburst.   In Fig. 1 we compare the 
observed spectrum with a model spectrum  calculated with the population 
synthesis code STARBURST99 (Leitherer \etal 1999). We have assumed a 
constant rate of star formation lasting for $t_*=10^{8}\,\yr$, a Salpeter 
initial mass function (IMF) from 1 to 100 $\msun$, and solar  metallicity. 
The spectrum is dominated by stars in the mass range $15-60\,\msun$, and 
has been convolved with the model of the opacity distribution of the \op 
forest clouds and Lyman-limit systems of Madau (1995). 
To improve the fit to the data a modest amount of reddening was introduced, 
$E_{B-V}=0.03$ for a SMC (absorption only, no scattering) extinction curve 
(Pei 1992). 

The agreement between the model spectrum and the data is rather good, 
although the observed flux shortward of 1 Ryd appears somewhat higher than
in the model. The discrepancy is, however, within the uncertainties in the 
spectral synthesis and intergalacitc opacity modeling. 
For comparison, we also show the 
spectrum of an instantaneous starburst after $10^{7}\yr$. This has 
basically no flux shortward of 1 Ryd. Fading starbursts with short duty 
cycles may thus be problematic (e.g. Kolatt \etal 1999). Note that there 
is no hint of a significant \hi\  photoelectric optical depth in the
local interstellar medium (ISM).

\begin{figurehere}
\epsscale{1.0} 
\plotone{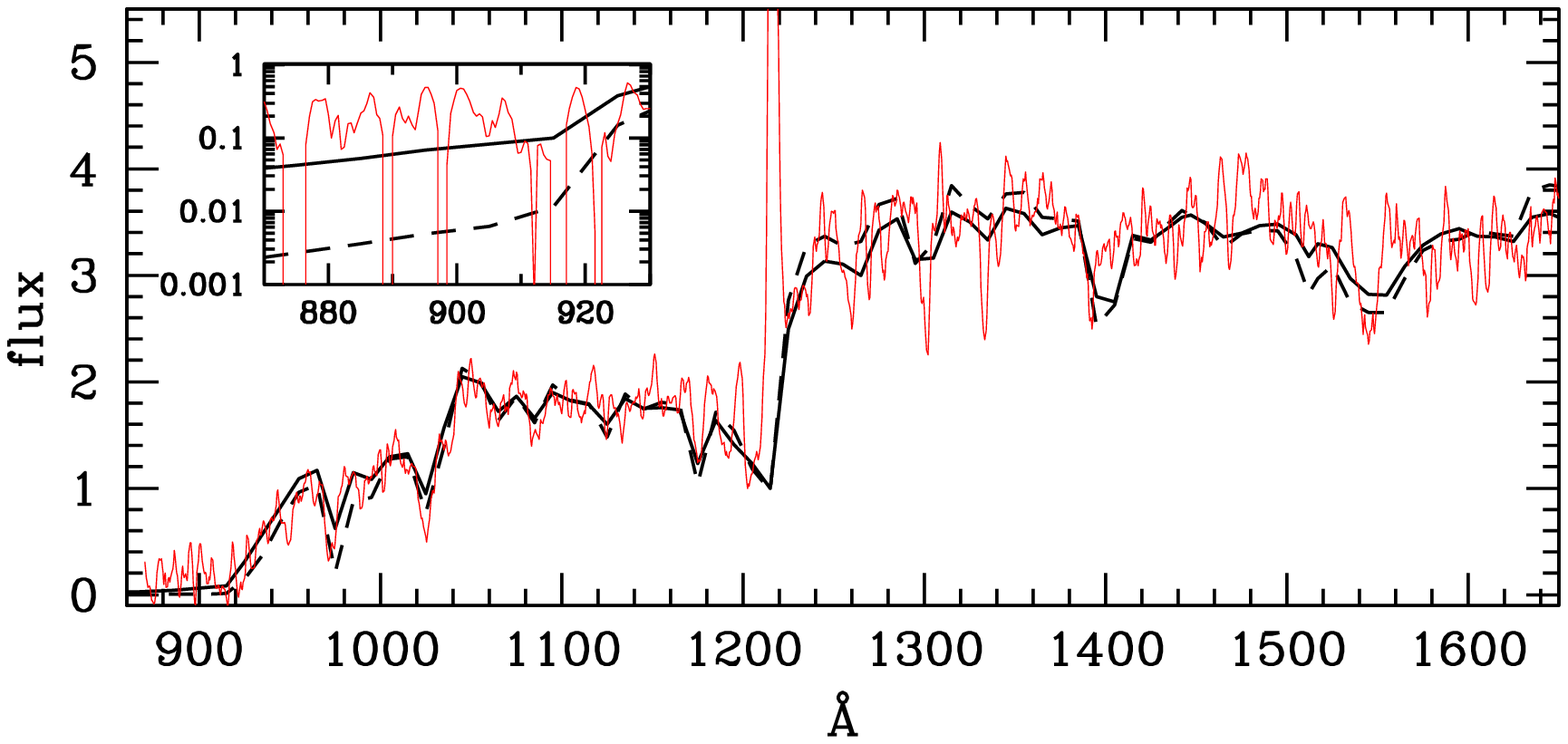}
\caption{\footnotesize {\it Thin solid curve:} composite spectrum of 29 
LBGs at $\langle z\rangle=3.4$ from Steidel \etal (2000). {\it Thick solid 
curve:} synthetic spectrum of a stellar population with continuous 
star formation and Salpeter IMF.  A modest amoung of reddening with a SMC 
extinction curve and $E_{B-V}= 0.03$ was included. {\it Dashed curve:} 
synthetic spectrum of an instantaneous starburst after $10^7 \yr$. 
The model spectra were calculated with STARBURST99 (Leitherer \etal 1999).} 
\end{figurehere}

\section{The neutral  hydrogen distribution and escape fraction
of Lyman-break galaxies.}  

If star formation in Lyman-break galaxies is indeed continuous and
lasts for about a hundred Myr there should be a substantial amount 
of neutral hydrogen present in these systems. The LBGs in  
Steidel \etal sample  have typical star  formation rates of 
$\dot M_*=10-50\,\msun \yr^{-1}$. 
They should thus contain  $M\approx 10^{9.5} (30\,\msun \yr ^{-1}/\dot M_*)
\;  (t_*/10^{8}\yr)\,\msun $  of cold gas if the cold gas to stellar 
mass ratio were of order unity. Taking a typical half-light radius 
of $2\kpc$ as characteristic size and assuming that the cold gas 
is distributed homogeneously, the typical density and column
density  should be $\sim 3 \,(M/10^{9.5}\msun)\, (r/2 \kpc)^{-3}\, \cm^{-3}$ 
and $2 \times 10^{22} (M/10^{9.5}\msun)  (r/2 \kpc)^{-2} \cm^{-2}$,
respectively. These values are higher than those of damped \lya systems 
and a large \hi optical depth would therefore be expected. 

A luminosity in ionizing photons of at least $\sim 10^{44.5}  
(M/10^{9.5}\msun) (r/2 \kpc )^{-3}\,\ergs$ is needed
to keep this large amount of gas photoionized. 
The required luminosity is a factor $10-30$ higher than that of the LBGs
in the sample. Lyman-break galaxies should
therefore either have a small cold/neutral gas mass to stellar mass   
ratio in their luminous regions or the cold gas must be contained  in small
compact regions with small covering factor.

\section{The UV escape fraction of galaxies and the redshift evolution
of the UV background}

The absence of a significant intrinsic \hi\ photoelectric 
opacity in the spectra of LBGs implies that those 
ionizing photons which avoid absorption by dust can escape the galaxy 
unimpeded. After correction for intergalactic 
absorption, the mean ratio of emergent flux density at (rest-frame) 1500 
\AA\ to 900 \AA\ is equal to $4.6\pm 1$ according to Steidel \etal (2000). 
Assuming this ratio is typical of all LBGs at $z\approx 3$ (this might
not be true since the composite spectrum is drawn from the
bluest quartile of the LBGs population) Steidel \etal
estimate $J_{21} = 1.2\pm 0.3$ for the galaxy contribution to the 
ionizing flux at the Lyman limit (here $J_{21}$ is measured in units of 
$10^{-21} \erg \s^{-1} \cm^{-2} \Hz^{1} \sr^{-1}$).
This is about a factor four larger than the contribution 
of QSOs at this redshift, and  may  solve the problem of 
the missing sources for H-reionization (Madau \etal 1999).

\begin{figurehere}
\epsscale{1.0}
\plotone{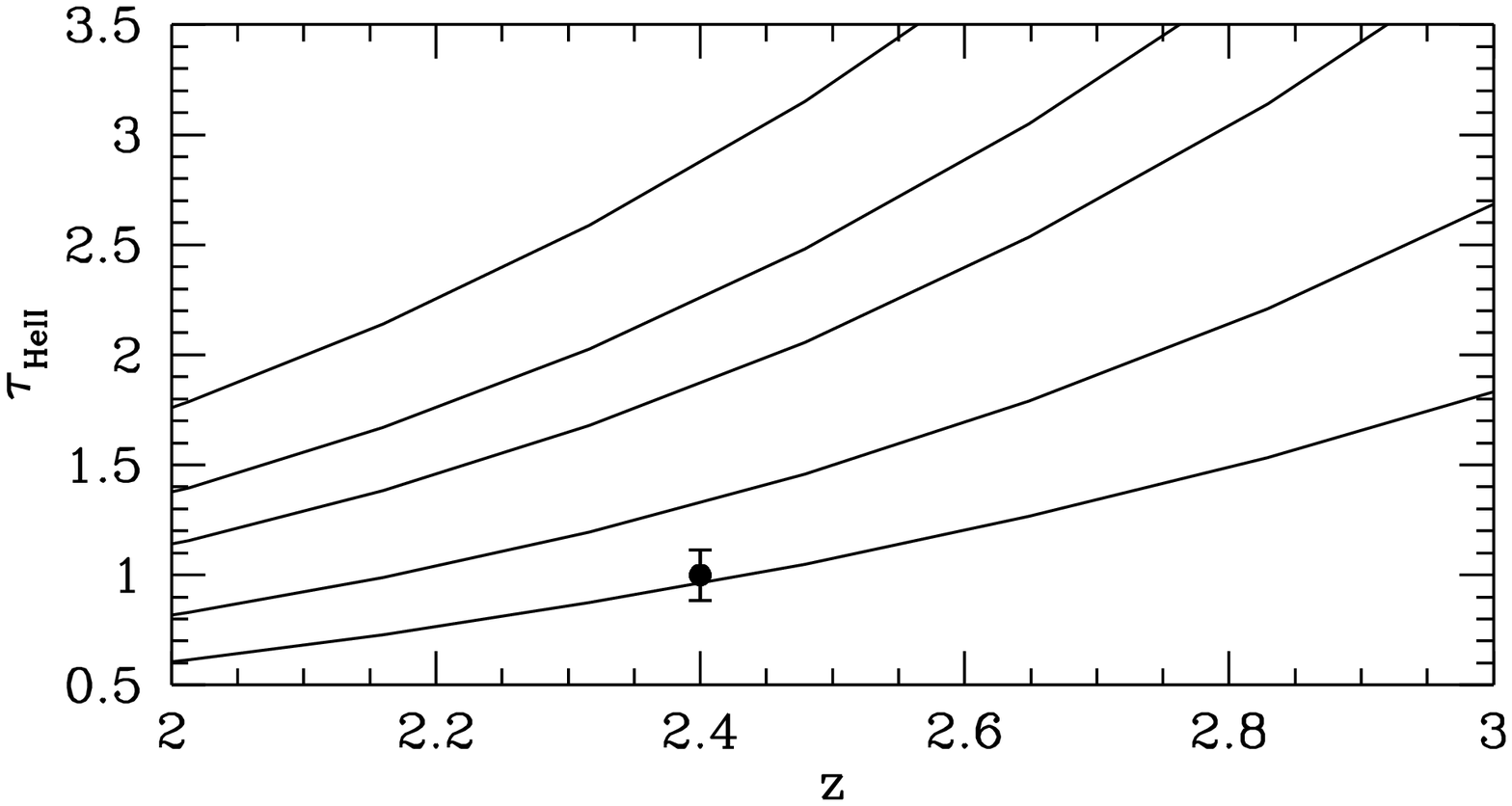}
\caption{\footnotesize  The \heii\ line blanketing opacity as a function 
of redshift 
for a UV background dominated by QSOs and stars. The different curves 
show the optical depth for a stellar contribution to the 
flux at 1 Ryd equal to (0, 0.5, 1.5, 2.5 and 5) times 
that of QSOs (from bottom  to top). The \hi\  
optical depth is fixed to $\tau_{\rm HI}=0.0038\,
(1+z)^{3.46}$, while the ionizing flux at $z=2.4$ from QSOs {\it only}
is $J_{21}=0.3$ in all cases (Haardt \& Madau 1996). 
The curves correspond to an escape fraction of Lyman-continuum photons 
from star-forming galaxies of (0, 10, 30, 50 and  100) percent.
The data point is from Davidsen et al. (1996).}
\end{figurehere}
\vspace{0.2cm}

{\it Could the UV background be dominated by stars at all redshifts?}
Comparison of both \hi\ and \heii\ absorption in quasar spectra can be
used to constrain the relative contribution of massive stars and QSOs to 
the metagalactic flux.  While in three out of four of the observed QSO 
spectra with a \heii\ \op
forest the interpretation is made difficult by the large fluctuations 
measured in the \heii\ scattering opacity -- perhaps the signature of a 
delayed $\heii\rightarrow \heiii$ reionization at $z\gta 3$ -- (see, e.g.,
Reimers \etal 1997; Heap \etal 2000) the spectrum  of HS1700+64 at 
somewhat lower redshifts ($2.4<z<2.7$; Davidsen \etal 1996) 
shows smaller fluctuations and allows a more reliable  
determination of the average \heii\ optical depth, $\overline \tau_{\rm 
HeII}=1.0\pm 0.07$. This value is in good agreement with that expected 
for a quasar-dominated ionizing background filtered by the 
IGM (Haardt \& Madau 1996) with a frequency dependence 
of the intrinsic quasar spectrum, $f_{\nu} \propto \nu^{-1.8}$ 
(Machacek \etal 2000; Theuns \etal 1998; Croft \etal 1997). 
In Fig. 2 we show the \heii\ line blanketing optical depth for a UV
background with different ratios of QSO and galaxy contribution, for a 
given \hi\ scattering opacity. The five curves correspond to an escape 
fraction of H-ionizing photons from galaxies of (0, 10, 30, 50, 100) percent
(bottom to top). For fixed \hi\ scattering opacity the \heii\ optical
depth increases with increasing stellar contribution to the total flux.

A UV background dominated by stars 
is inconsistent with the observed ratio of \heii\ and \hi optical 
depth at $z\sim 2.4$. The average escape fraction of LBGs at this 
redshift must be $\la 10 \%$ similar to the small $f_{\rm esc}$ -- 
between 5 and 10\% -- inferred in local starbursts 
(Leitherer \etal 1995; Hurwitz \etal 1997)
but very different  from the close to $100 \%$ of the Steidel et al 
subsample at $z\sim 3.4$. If the results of Steidel \etal were representative 
of the entire LBG population, the average escape fraction would have to 
decrease very rapidly with decreasing redshift.

\begin{figurehere}
\epsscale{1.0}
\plotone{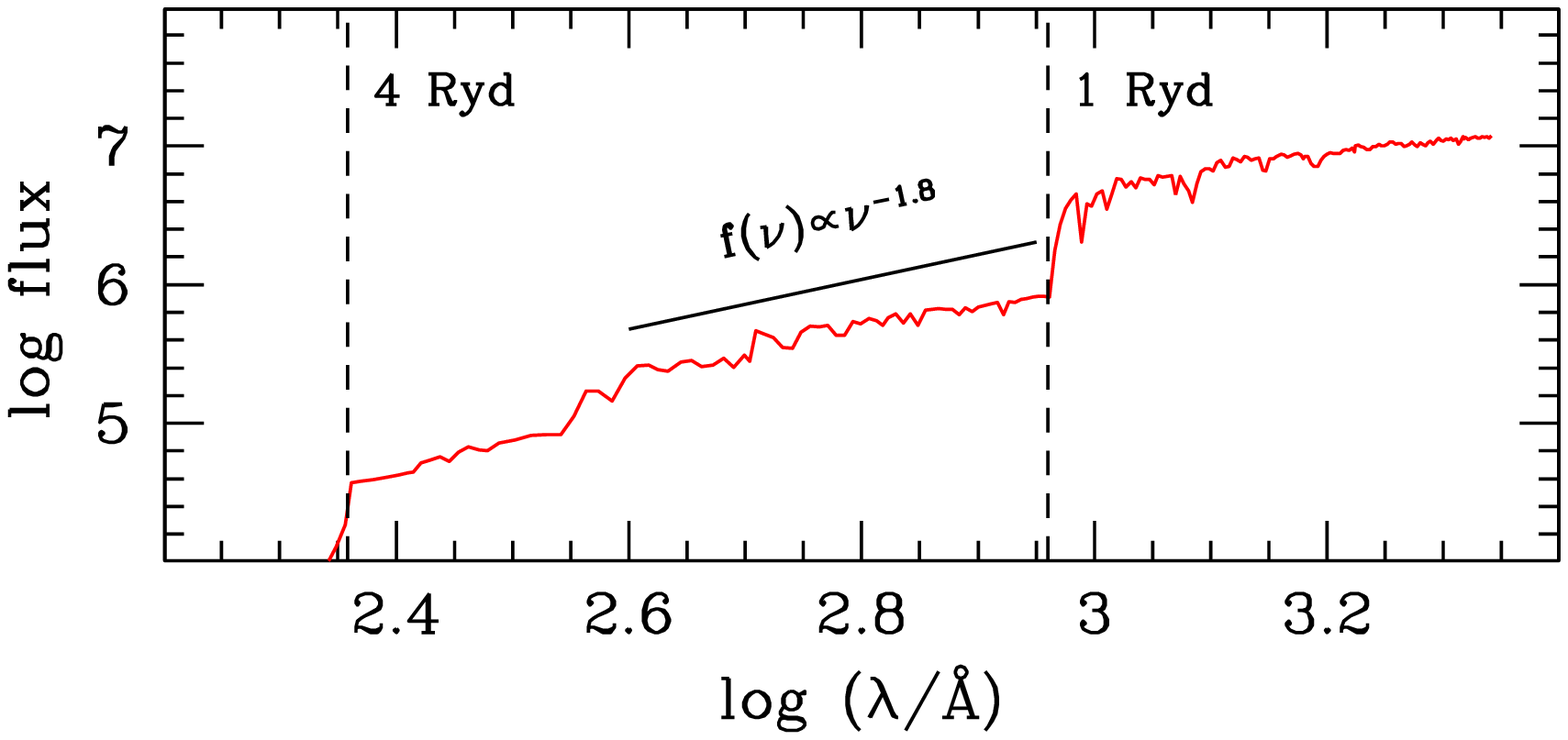}
\caption{\footnotesize  Extrapolated far-UV synthetic spectrum of a 
stellar population with continuous star formation. A SMC reddening curve 
as modeled by Pei (1992) and the dust opacity distribution of LBGs as 
given by Steidel \etal (1999) have been assumed. The line indicates the 
slope of a typical QSO spectrum with $f_{\nu} \propto \nu^{-1.8}$.} 
\end{figurehere}

\section{A UV background dominated by stellar  sources at high redshift}

Having found a simple model which reproduces the observed spectrum 
in the wavelength range $870\,$\AA\ to $1600\,$\AA\ we can now extrapolate 
the UV spectrum to shorter wavelength. The result is shown in Fig.
3. We have assumed  the full range of dust optical
depth as reported by Steidel \etal (1999) with  no dependence of 
dust optical depth on galaxy luminosity, together with a SMC 
reddening curve (absorption only, no scattering; Pei 1992). We also 
assumed that the 
null detection of \hi\  absorption due to the ISM in the Steidel 
\etal sample is representative of all LBGs. 
Just beyond the Lyman limit the spectrum is very similar to that of a 
typical QSO, somewhat harder than the canonical
$-1.8$ power law (Zheng \etal 1998). It is only at wavelengths shorter 
than $400\,$\AA\  that it softens significantly compared to a quasar
spectrum. Beyond the \heii\ Lyman edge at 4 Ryd
there is a strong break.

The similarity  to a QSO spectrum just 
beyond 1 Ryd may explain the already  high temperatures of $10^4 \K$ 
at $z>3$ (Haehnelt \& Steinmetz 1998; Schaye \etal 2000).
A UV background which is progressively  dominated by O stars may 
also be the solution to the increasing  \SiIV/\CIV\ ratio with 
increasing redshift observed in metal absorption systems of intermediate 
column densities (Songaila \& Cowie 1996; 
Giroux \& Shull 1997; Boksenberg \etal 1998).

{\it How different could the integrated stellar spectrum be?}
The upper cut-off of the IMF and the metallicity 
will strongly affect how far the spectrum extends into the 
far UV (Tumlinson \& Shull 2000; Bromm \etal 2000; 
Oh \etal 2000). 
We have assumed an upper cut-off of $100 \msun$ and solar metallicity. 
The dust opacity in the far UV depends on smaller grains 
than usually probed by reddening analysises and is rather uncertain. 
In the the model of Pei \etal (1992) for the SMC opacity used here, 
the dust opacity 
drops shortwards of $1000\,$\AA.  
This makes the spectrum  bluer not redder 
at wavelengths shorter than $1000\,$\AA. 
The stellar spectrum of massive stars with strong mass loss 
are  complex and require non-LTE calculations including the effect 
of stellar wind outflow and spherical extension. In particular, the strength 
of the 4 Ryd break depends significantly on  metallicity, 
mass loss rates and details of the modelling (Kudritzki 2000; Pauldrach et 
al. 2000).

\section{The baryon fraction of the Universe}

One interesting consequence of the  large amplitude of the UV 
background suggested by Steidel \etal is the implied conflict 
of the \op opacity measurements in QSO spectra with the 
nucleosynthesis constraint for the baryon density. The optical depth 
for \op scattering of the photoionized IGM scales as (Rauch \etal 1997a) 
$\tau_{\rm Ly\alpha}\propto H(z)^{-1} 
(\Omega_{\rm b} h^2)^2 T^{-0.7} \Gamma^{-1},$ 
where $\Gamma$ is the  photoionization rate, $T$ is 
the gas temperature, $H(z)$ is the Hubble constant at redshift 
$z$, $\Omega_{\rm b}$ 
is the baryon density parameter, and $h$ is 
the present-day Hubble constant in units of $100\kms\,\mpc^{-1}$. 
The investigation  of Rauch \etal  implies that for a 
nucleosynthetic value of $\Omega_{\rm b} h^2 =0.019$ (Burles \& Tytler 1998) 
the \op opacity can be matched if 
$\Gamma \sim 5 \times  10^{-13} \s^{-1}$. For a spectrum $f_{\nu} \propto
\nu^{-1.8}$ this  corresponds to an ionizing flux $J_{21} =0.12$.
The large value of $J_{21}=1.2\pm 0.3$ suggested  
by Steidel \etal implies instead $\Omega_{\rm b}h^2\approx 0.06$. The 
D/H measurements are therefore in strong conflict 
with the \op opacity measurements if both the standard nucleosynthesis
calculations and the value of  $J_{21}$ suggested by 
Steidel \etal are correct. This discrepancy may be even 
stronger if there were a significant contribution to the 
H-ionizing background from faint \op emitters with weak stellar 
continuum flux as suggested by Kudritzki et al. (2000).

In Fig. 4 we have compiled from the literature a number of  values 
for $\Omega_{\rm b}h^2$ implied by the measured  \op opacity 
assuming $J_{21}= 0.6$ as may be appropriate if  
LBGs with redder spectra than those in the 
Steidel et al. sample had significant smaller 
escape fraction for H-ionizing photons.
The values at $z=3$ are taken from  Rauch \etal (1997b), Gnedin (1998), 
Nusser \& Haehnelt (2000), Theuns \etal (2000), and at 
$z=2$ from Cen \etal (1994) and Hernquist \etal (1996). 
We have scaled all values to a temperature of 15000 K at $z=3$ and 11000 K 
at $z=2$ as suggested by the temperature determination of the IGM
by Schaye \etal (2000). All values for $\Omega_{\rm b}$  scale 
$\propto J_{21}^{0.5}$ and $\propto T^{0.35}$.  
We also show the nucleosynthesis constraint from 
the D/H measurement as given by Burles \& Tytler (1998) and the
combined constraints from the CMB experiments BOOMERANG and MAXIMA 
(Jaffe \etal 2000).  

\begin{figurehere}
\epsscale{1.0}
\plotone{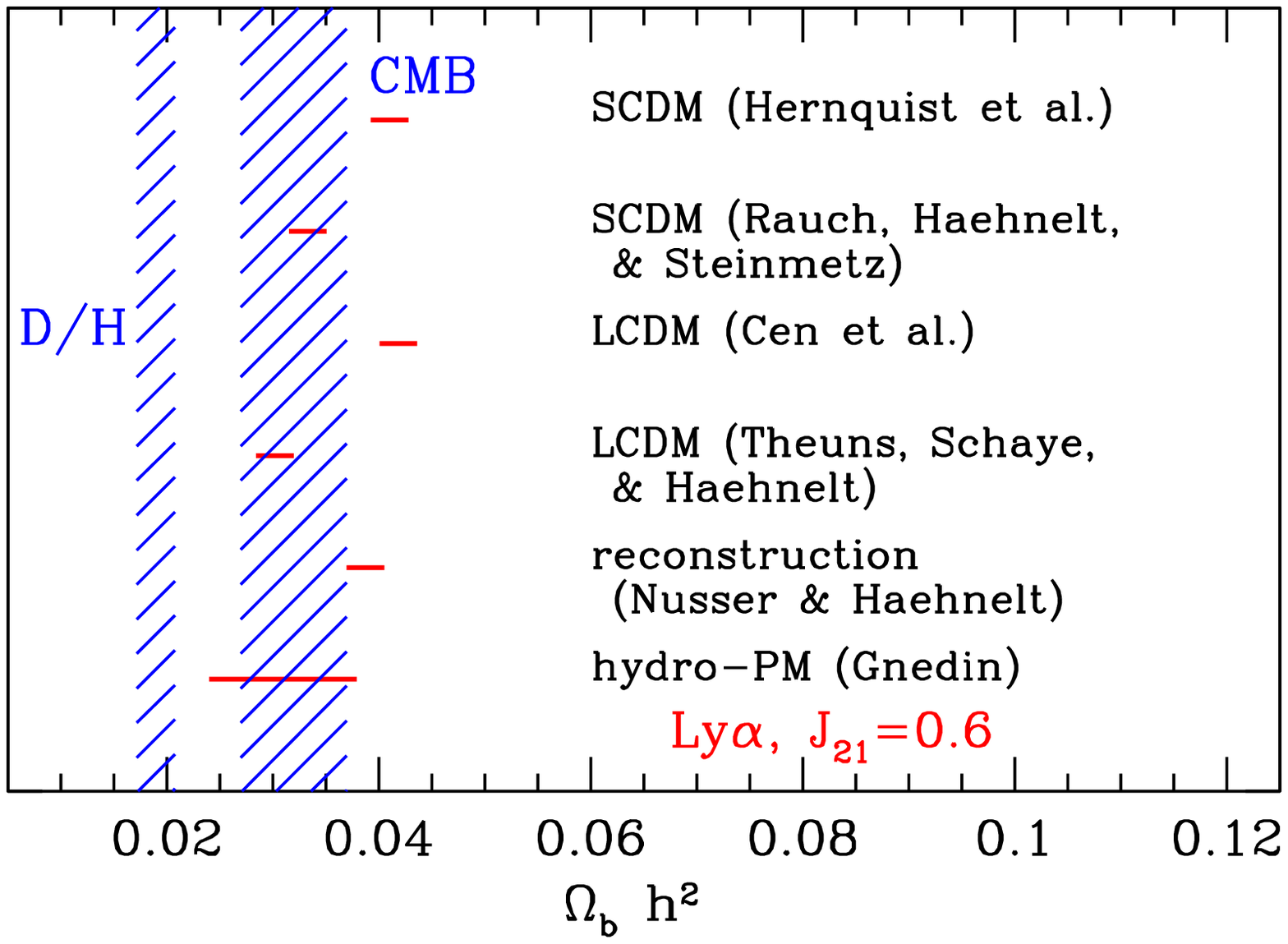}
\caption{\footnotesize  The baryon fraction inferred from the \op
opacity as described in the text. The values inferred from 
D/H measurements (Burles \& Tytler 1998)  and CMB experiments 
(Jaffe \etal 2000) are also shown.} 
\end{figurehere}

\section{Discussion and Conclusions}

We have shown that the detection of flux beyond 
the Lyman limit from LBGs is most easily explained as being 
due to the O stars present in a star forming region with ongoing 
star formation, a Salpeter IMF, modest or negligible reddening, and 
negligible absorption due to \hi\  in the local ISM. 
Fading starbursts in which star formation has ceased   
for $10^{7} \yr$  or more cannot reproduce the observed flux 
shortward of 1 Ryd.
The small \hi\  optical depth cannot  be due to photoionization.
The cold neutral gas from which they stars form must be contained in 
compact small regions with small covering factor. 

The resulting ionizing background will be
dominated by O stars shortward of 
the Lyman limit and is as hard as a QSO spectrum between 1 and 3
Ryd. Shortward of 3 Ryd the spectrum is softer than a 
QSO spectrum and there is strong drop shortward
of 4 Ryd. 
A consistent picture seems to emerge  where the hydrogen of the 
Universe is reionized by stars at $z>6$ while the reionization 
of helium is due to QSOs and is delayed to $z\sim$ 4 or 3. The 
contribution of stars to the ionizing UV background is 
equal or larger than that  QSOs at least up to $z\sim 3$. At smaller 
redshift  the ratio of \heii\  and \hi\  opacity and the 
observed \SiIV/\CIV\ ratios in the associated metal absorption 
systems of intermediate column density QSO absorption systems argue for 
a UV background dominated by QSOs. This requires a substantial  
decrease of the escape fraction of ionizing photons 
from star-forming galaxies with decreasing redshift. 
The amplitude of the UV background inferred from the composite
spectrum of Steidel \etal makes the  measurements of the opacity 
of the \op forest in QSO absorption spectra inconsistent 
with D/H measurements if the spectrum is representative 
for all LBGs and nucleosynthesis calculations 
for D/H are correct.  The inferred value for $\Omega_{\rm b}$
is similar to that that indicated  by the CMB experiments 
Boomerang and Maxima if $J_{21}= 0.4-0.6$.

\acknowledgments
We are grateful to C. Steidel for providing the composite 
spectrum. We thank S. Charlot, G. Efstathiou, M. Pettini 
and S. White for helpful discussions and comments.
This work was supported by the EC TMR network 
``The Formation and Evolution of Galaxies'', 
by the EC RTN network ``The Physics of the Intergalactic 
Medium'', and  by NASA through ATP grant NAG5-4236 and grant 
AR-06337.10-94A from the Space Telescope Science Institute (P.M.).


\end{document}